\documentclass{article}
\usepackage{graphicx}
\usepackage{svg}
\usepackage[T1]{fontenc}
\usepackage{amsmath,amssymb}
\usepackage[utf8]{inputenc}
\usepackage{ismir}
\usepackage{amsmath,cite,url}
\usepackage{graphicx}
\usepackage{color}
\usepackage{tabularx}

\usepackage{multirow}
\usepackage{url}
\usepackage{caption} 

\usepackage{booktabs} 

\usepackage{threeparttable} 

\usepackage{siunitx}
\newcommand{\best}[1]{\textbf{#1}}  % 粗體小工具

\newcolumntype{Y}{S[table-format=1.3,parse-numbers=false]} % 數值+括號一起放
\usepackage{array}
\usepackage[table]{xcolor}
\definecolor{wincol}{RGB}{205,230,255}   % 淺藍（贏）
\definecolor{losecol}{RGB}{255,224,224}  % 淺紅（輸）

\newcolumntype{B}{@{\hspace{0.5em}}!{\vrule width 0.85pt}@{\hspace{0.5em}}}

\title{Separate-and-Detect: Unified Drum Transcription and Stem Generation via Latent Diffusion}

\multauthor
  {Wei-Han Hsu$^{1,2}$, Chih-Cheng Chang$^2$, Bo-Yu Chen$^3$, Li Su$^2$, Yi-Hsuan Yang$^{4}$}
  {$^1$Data Science Degree Program, National Taiwan University and Academia Sinica, Taipei, Taiwan\\
   $^2$Institute of Information Science, Academia Sinica, Taiwan\\
   $^3$Rhythm Culture Corporation \\
   $^4$AI Center of Research Excellence (AI-CoRE), National Taiwan University\\
   {\tt\small ddman821101@gmail.com, affige@gmail.com, lisu@iis.sinica.edu.tw} % 請填入真實 Email
  }

\def\authorname{W.-H. Hsu, C.-C. Chang, B.-Y. Chen, L. Su, and Y.-H. Yang}

\begin{document}

\maketitle

\begin{abstract}
Automatic Drum Transcription (ADT) is commonly formulated as a direct mapping from a music mixture to symbolic drum events. While effective for transcription, this formulation discards the acoustic stems that are useful for editing, remixing, and production. We revisit an alternative separate-and-detect formulation, where a drum source separation front end first produces five editable drum stems, and a fixed onset detector then converts each stem into symbolic events. The separator is built on a five-stem latent diffusion model that jointly generates kick, snare, toms, hi-hats, and cymbals in a compact VAE latent space. We further study two training-only auxiliary branches---an onset branch (OB) and a timbre branch (TB)---which shape the separator during learning but are discarded at inference. Trained on synthetic drum multitracks and evaluated on MDB Drums and ENST-Drums, the proposed pipeline consistently improves over a strong U-Net-based drum separation baseline in overall transcription F1. It also outperforms a representative end-to-end ADT system on kick and snare F1 under our evaluation protocol, while additionally providing separated audio stems. The ablation results show that OB gives the most stable transcription gains, whereas TB changes the trade-off between reconstruction, acoustic stem quality, and onset detection. These results suggest that generative drum demixing can serve not only as a source separation model, but also as a practical front end for interpretable drum transcription.
\end{abstract}

\section{Introduction}
\label{sec:intro}

Automatic Drum Transcription (ADT) aims to convert audio into symbolic drum events such as kick, snare, tom, hi-hat, and cymbal hits. The task is especially difficult in full-mixture music, where drum events overlap with vocals, harmonic instruments, and other sources. Recent deep learning systems have improved ADT performance, but accurate transcription from complex mixtures remains challenging \cite{adtof}.

Most recent ADT systems approach this problem as direct event prediction from the mixture. %This design is compact, but it also collapses the task into a single symbolic output and discards the underlying drum audio. 
While direct event prediction is highly efficient, it only outputs symbolic events and cannot retrieve the underlying, editable audio waveforms of individual drums.
In this work, we revisit a separation-based alternative. We first obtain a drum-dominant signal from the music mixture, then decompose it into five drum %-piece 
stems, and %finally 
apply onset detection to each stem. This separate-and-detect formulation makes the intermediate representation explicit: the system produces both symbolic drum events and editable audio stems. Such stems are useful in music production settings, where drum parts are not only transcribed, but also edited, replaced, layered, or remixed. In many contemporary production workflows, kick and snare form the core of the groove: the kick often defines the low-frequency pulse and rhythmic drive, while the snare marks the backbeat and strongly shapes the perceived feel of the drum pattern. For drummers, beat makers, and producers, reliable access to these two components is %therefore 
especially valuable, since small timing or timbral changes to kick and snare can alter the character of the entire rhythm section.

To implement this, we build a five-stem drum source separation model based on latent diffusion \cite{msg-ld}. The model operates in a compact Variational Autoencoder (VAE) latent space and jointly predicts kick, snare, toms, hi-hats, and cymbals. The separated stems are rendered back to audio with a neural vocoder \cite{musicldm}, and each stem is then processed by a fixed onset detector. We also investigate two training-only auxiliary branches: an onset branch (OB) that encourages sparse percussive structure, and a timbre branch (TB) based on DrumGAN descriptors \cite{timbre}. These branches are used only to shape the separator during training and are removed from the inference path.

This paper makes three contributions. First, we take full-mixture drum transcription as a separate-and-detect pipeline and show that per-stem onset detection can be a practical alternative to direct end-to-end ADT. Second, we introduce a five-stem latent diffusion separator for drum transcription and show that it improves over a strong drum separation baseline in downstream transcription accuracy. Third, we analyze how the training-only OB and TB auxiliary branches affect the trade-off between reconstruction quality, acoustic stem quality, and transcription performance. The code and demos are publicly available online.\footnote{\urlstyle{same}
\raggedright Code: \url{https://github.com/ddman1101/Separate-and-detect};
\\ 
\raggedright demo page: \url{https://ddman1101.github.io/Separate-and-Detect-demo/}.
}

\begin{figure*}[!t]
    \centering
    % \includesvg[width=0.95\textwidth]{SnD_arch_v2_8}
    \includegraphics[width=0.95\textwidth]{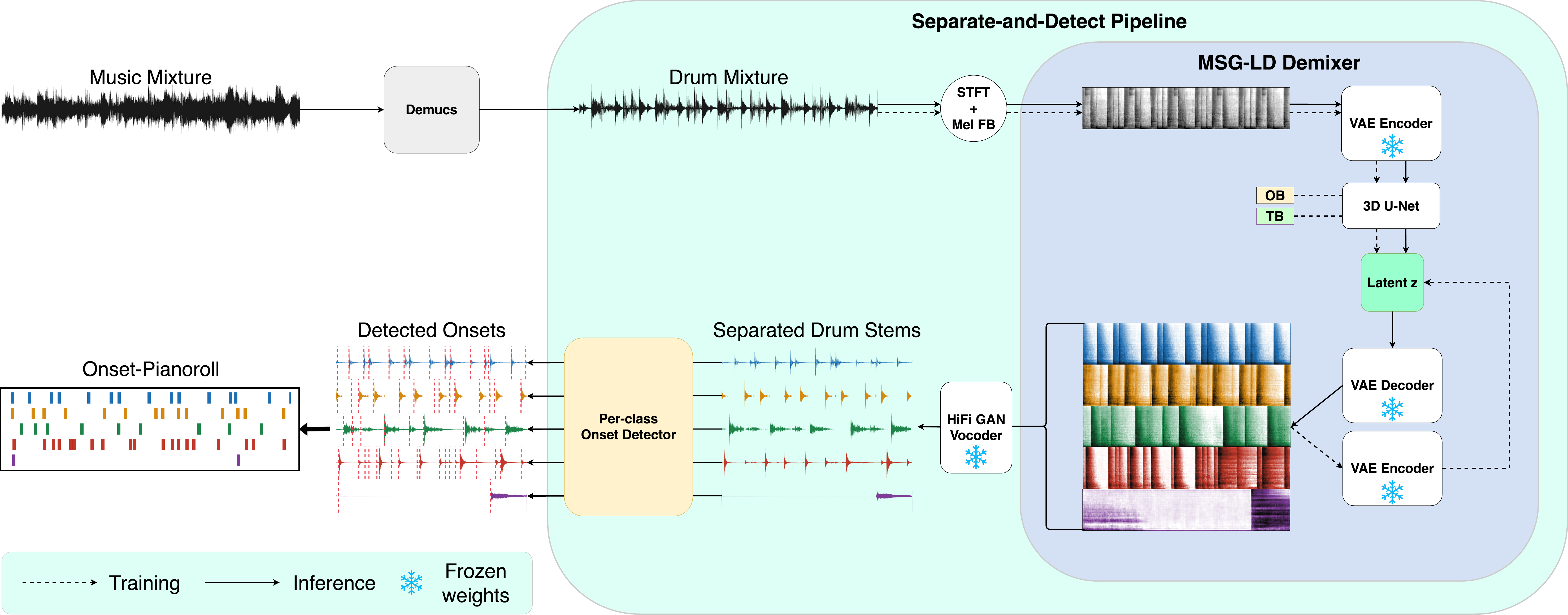}
    \caption{Overview of the proposed separate-and-detect pipeline. Given an input music mixture, we first obtain a drum mixture using Demucs, and then apply the MSG-LD demixer to generate separated drum stems. A per-class onset detector subsequently converts the separated stems into symbolic drum onsets, represented as a pianoroll. Solid and dashed arrows denote the data flow during inference and training, respectively, while the snowflake symbol indicates components with frozen weights. OB and TB denote the onset and timbre branches, respectively.}
    \label{fig:pipeline}
\end{figure*}

\section{Related work}\label{sec:related_work}

\subsection{Diffusion Models for Music Source Separation}
Diffusion models have been explored for music source separation (MSS) by learning joint multitrack distributions that support both separation and generation~\cite{MSDM}. \emph{Latent} variants denoise in a compact VAE latent space and render audio with a neural vocoder~\cite{msg-ld,musicldm}; user/text-guided demixing further relaxes fixed four-stem assumptions~\cite{guidesep_2025}. Prior diffusion work typically targets generic 4-stem or instrument-agnostic settings. We instead specialize a latent diffusion separator to the five drum sub-stems (kick, snare, toms, hi-hats, cymbals) and exploit the conditioning interface to attach \emph{training-only} onset/timbre heads that shape separator features with no test-time overhead.

\subsection{Drum Source Separation}
General-purpose MSS backbones include waveform or hybrid spectrogram-waveform models such as HT-Demucs, Open-Unmix, and band-split RNN, with Spleeter widely used for 4-stem MSS~\cite{htdemucs,stoter2019open,spleeter,band-split-rnn}. For five-stem DSS, \emph{StemGMD} standardizes the protocol \cite{larsnet_stemgmd}, and \emph{LarsNet} trains stem-specific U-Nets as a representative baseline of stage~2~\cite{larsnet_stemgmd}. Recent benchmarking adapts general demixers to drums and analyzes time/TF/hybrid trade-offs~\cite{ddsp}. In our pipeline, HT-Demucs is used only at inference to obtain a drum-dominant signal from full mixtures; the proposed latent diffusion model is trained and evaluated as the five-way drum sub-stem separator.

\subsection{Drum Transcription}
Early ADT relied on hand-crafted onset detectors combined with non-negative matrix factorization (NMF) and rule-based heuristics~\cite{boeck_superflux,boeck_rnn}. Modern approaches use neural models for drum event prediction from audio, including convolutional-recurrent architectures and end-to-end systems. Large-scale audio-to-MIDI aligned training data have also been explored to improve automatic drum transcription~\cite{wei2021improving}, with ADTOF as a representative strong baseline~\cite{vogl2017drum,adtof}. Very recent work has also begun to cast ADT as a generative problem: \emph{Noise-to-Notes} formulates drum transcription as conditional diffusion generation of drum events and velocities~\cite{noise_to_notes}. 

A different recent line explores the intersection of transcription and source separation. %\emph{Enhanced Automatic Drum Transcription via Drum Stem Source Separation} augments 
Recently, Riley and Dixon ~\cite{enhanced_adt_sep} augmented ADTOF with drum stem separation to expand the output taxonomy and estimate MIDI velocities, aiming for more realistic symbolic drum tracks. Similarly, the \emph{Inverse Drum Machine}~\cite{inverse_drum} leverages an analysis-by-synthesis framework to jointly optimize drum transcription and one-shot sample synthesis, achieving drum source separation using only transcription annotations for training.

Unlike these joint or refinement-based approaches, our work explores a sequential ``separate-and-detect'' pipeline as the primary inference path. Rather than directly predicting events from the full mixture, relying on joint synthesis or using separation only as a downstream refinement step, we first decompose the music into drum sub-stems and then apply a frozen onset detector per stem. This design allows us to study drum source separation not only as an audio task but also as a front end optimized for transcription utility.

\begin{figure}[t]
    \centering
    \includegraphics[width=\columnwidth]{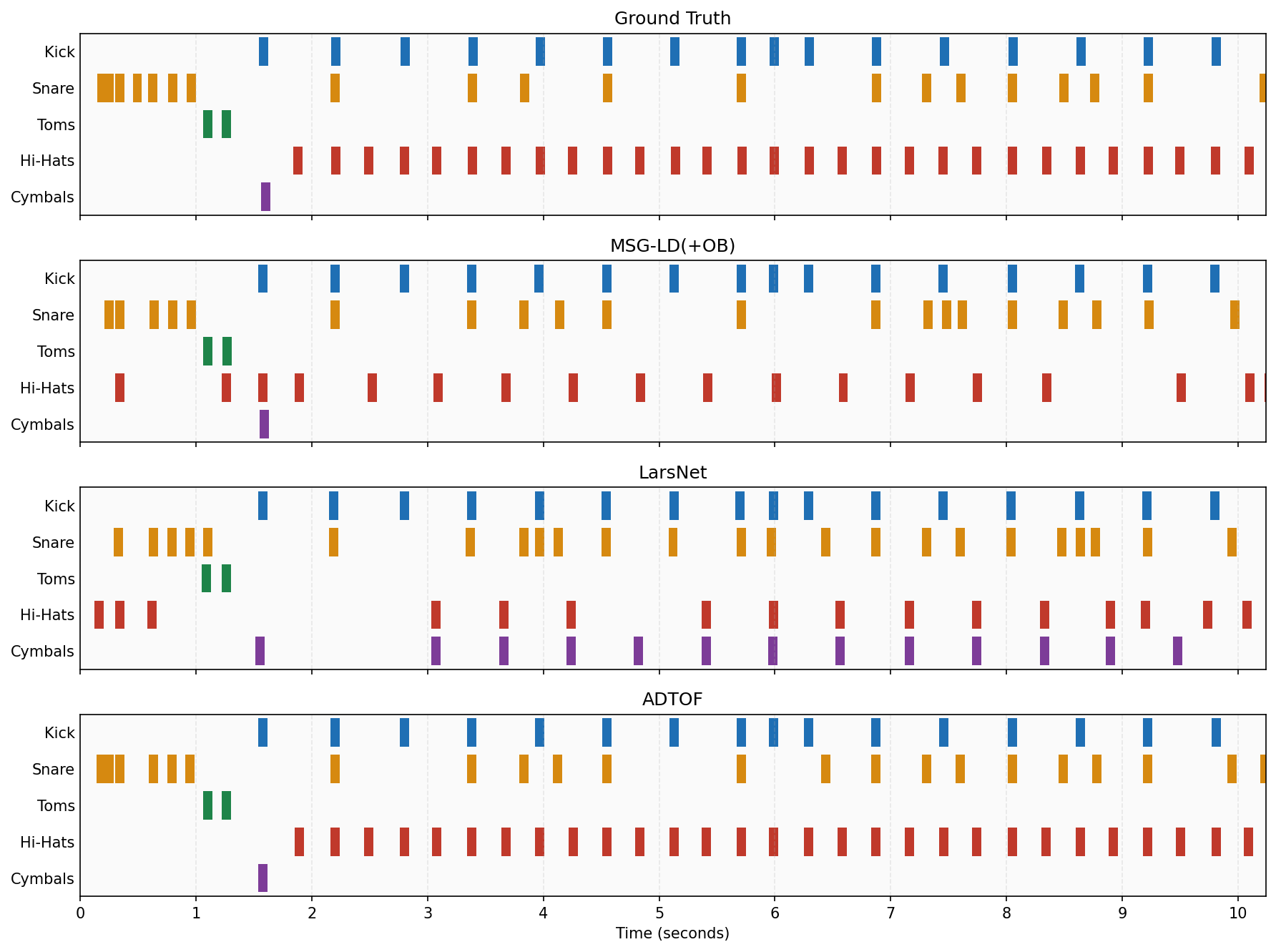}
    \caption{A pianoroll comparison on an unseen test excerpt. MSG-LD (+OB) preserves the kick/snare rhythmic backbone and avoids several spurious events produced by the separation-based LarsNet baseline. At the same time, ADTOF remains strong on dense hi-hat activity, illustrating the continuing challenge of broadband drum textures for separate-and-detect systems.}
    \label{fig:pianoroll}
\end{figure}

\begin{figure}[t]
    \centering
    \includegraphics[width=\columnwidth]{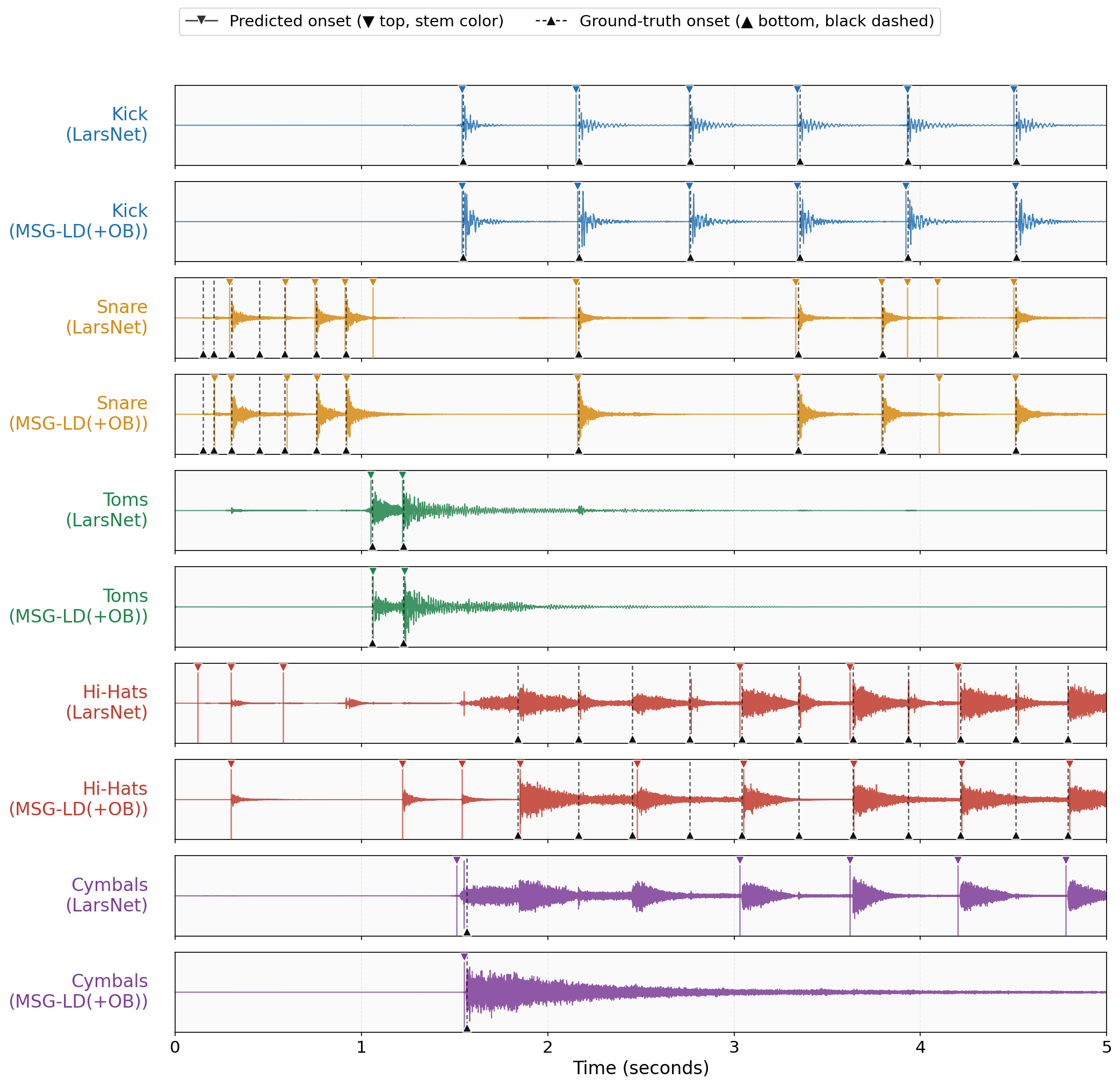} 
    \caption{Acoustic analysis of the separated stems for a 5-second excerpt corresponding to the pianoroll in Figure~\ref{fig:pianoroll}. The top markers (colored triangles) denote predicted onsets, while the bottom markers (black dashed lines) indicate ground-truth events. Compared with LarsNet, MSG-LD (+OB) yields cleaner kick and snare transients and reduces cross-stem leakage in silent regions, supporting more reliable onset detection. The hi-hat and cymbal tracks also illustrate the remaining difficulty of dense broadband drum textures.}
    \label{fig:waveform_comparison}
\end{figure}

\section{METHODS}
\label{sec:method}

We formulate drum transcription as a sequential separate-and-detect pipeline. Given an input drum mixture, the system first isolates it into five drum-piece stems corresponding to kick, snare, toms, hi-hats, and cymbals. Each separated audio stem is then processed by an independent, fixed onset detector to obtain symbolic events, which are finally merged into a multi-track drum pianoroll. 

The primary engine of our pipeline is a specialized five-stem \emph{multi-track latent diffusion} separator. The model is trained on drum mixtures paired with isolated multitrack stems. During full-mixture inference, a fixed music source separation (MSS) front end is first employed to extract a drum-dominant signal, which is then passed to our proposed separator. Notably, neither the MSS front end nor the onset detection backend is involved in the separator's training phase. This design isolates the five-stem drum sub-separation task from variations introduced by the upstream full-mixture separator, while still enabling full-mixture inference through a fixed Demucs front end.

\subsection{Multi--track Latent Diffusion Separator}

Our separator is built upon the MSG-LD framework \cite{msg-ld}. We retain the core VAE, U-Net denoiser, and HiFi-GAN vocoder \cite{hifigan} architectures while adapting the track conditioning to accommodate the $K=5$ drum sub-stems. Let \(x\) be the drum mixture and \(\{y_k\}_{k=1}^{K}\) the isolated stems; their mel spectrograms are encoded into a compact VAE latent space. The mixture latent $\mathbf{c}$ serves as the conditioning signal for the diffusion process.

The denoising objective follows the standard noise-prediction formulation:
\begin{equation}
\mathcal{L}_{\mathrm{DDPM}} = \mathbb{E}_{\mathbf{z},t, \boldsymbol{\epsilon}} \left[ \left\| \boldsymbol{\epsilon} - \epsilon_{\theta}(\mathbf{z}_t,t,\mathbf{c}) \right\|_2^2 \right],
\end{equation}
where \(\mathbf{z}_t\) is the noisy target stem latent, \(\boldsymbol{\epsilon}\) is the sampled Gaussian noise, and \(\epsilon_{\theta}\) is the U-Net. The conditioning $\mathbf{c}$ includes the mixture latent and five learnable track tokens. After denoising, the VAE decoder converts the predicted stem latents into mel spectrograms \(\{\widehat{\mathbf{m}}_k\}_{k=1}^{K}\), which the HiFi-GAN vocoder then renders into waveform signals. We keep the HiFi-GAN decoder from MSG-LD to preserve the original backbone and isolate the effects of drum-specific conditioning and auxiliary objectives.

\subsection{Auxiliary Objectives}

To guide the separator's representation learning without increasing its inference complexity, we introduce two optional auxiliary branches attached to the penultimate decoder feature map \(\mathbf{h}\) of the U-Net. These branches are utilized solely during training and are discarded at test time.

\vspace{1mm}\noindent\textbf{Onset branch (+OB). } 
This branch predicts coarse stem-wise onset maps $\widehat{\mathbf{Y}}_{\mathrm{on}}$ from \(\mathbf{h}\) using a $1\times1\times1$ 3D convolution and lightweight upsampling blocks. By collapsing the frequency axis into frame-level onset logits, it encourages the model to prioritize the impulsive energy envelopes essential for transcription.

\vspace{1mm}\noindent\textbf{Timbre branch (+TB). } 
The timbre branch provides global acoustic regularization. It applies global average pooling to \(\mathbf{h}\), concatenates it with stem embeddings, and utilizes an MLP to predict 7-D DrumGAN descriptors \cite{timbre} \(\widehat{\mathbf{Y}}_{\mathrm{tim}} \in [0,1]^{K \times 7}\). These encompass \textit{brightness}, \textit{hardness}, \textit{depth}, \textit{roughness}, \textit{boominess}, \textit{warmth}, and \textit{sharpness}. 

\vspace{1mm}\noindent\textbf{Overall objective. } 
The separator is optimized through a multi-task objective defined as:
\begin{equation}
\mathcal{L} = \mathcal{L}_{\mathrm{DDPM}} + \lambda_{\mathrm{on}}\mathcal{L}_{\mathrm{onset}} + \lambda_{\mathrm{tim}}\mathcal{L}_{\mathrm{timbre}},
\end{equation}
where $\mathcal{L}_{\mathrm{onset}}$ is the Focal Binary Cross-Entropy (FocalBCE, $\alpha=0.25, \gamma=2.0$) between predicted and target onset maps, and $\mathcal{L}_{\mathrm{timbre}}$ is the $L_2$ loss for timbral descriptors. Auxiliary loss weights $\lambda$ are set to zero when a branch is disabled.

% \subsection{Onset-Based Transcription}

% At test time, the separated audio stems are independently processed by an off-the-shelf onset detection backend, specifically relying on the \texttt{madmom} CNN-based onset processor. For each drum class, we utilize a fixed detector configuration optimized once on a held-out development split. Notably, these class-specific configurations include customized signal pre-processing-such as targeted low-pass, high-pass, or band-pass filtering-applied prior to the CNN activation, alongside tailored peak-picking parameters including activation thresholds and minimum inter-onset intervals. 

% Once detected, the discrete events from the five individual stems are temporally aggregated to output the final transcription. Crucially, this external backend receives no gradient from the diffusion model, allowing us to strictly isolate the impact of the generative separation quality on the downstream symbolic analysis.

\subsection{Onset-Based Transcription}

% At test time, the separated audio stems are independently processed by an off-the-shelf onset detection backend, specifically relying on the \texttt{madmom} CNN-based onset processor. 
At test time, the separated audio stems are independently processed by an off-the-shelf onset detection backend. We apply onset detection to the final reconstructed waveforms so that transcription performance reflects whether the generated stems preserve the impulsive energy patterns needed for drum transcription. Specifically, we rely on the \texttt{madmom} CNN-based onset processor. For each drum class, we utilize a fixed detector configuration optimized once on a held-out development split using Optuna. Specifically, we tune the peak-picking parameters, yielding class-specific activation thresholds ranging from 0.35 to 0.60 and minimum inter-onset intervals from 30 to 50\,ms.

Once detected, the discrete events from the five individual stems are temporally aggregated to output the final transcription. Crucially, this external backend receives no gradients from the diffusion model, allowing us to strictly isolate the impact of the generative separation quality on downstream symbolic analysis.

\begin{table}[t]
\centering
\small
\setlength{\tabcolsep}{4pt}
\begin{tabular}{l c cccc}
\toprule
& & \multicolumn{4}{c}{\textbf{MSG-LD}} \\
\cmidrule(lr){3-6}
\textbf{Class} & \textbf{LarsNet} & \textbf{Vanilla} & \textbf{(+OB)} & \textbf{(+TB)} & \textbf{(+OB+TB)} \\
\midrule
Kick    & 4.57 & 1.13 & 1.03 & 0.84 & \textbf{0.78} \\
Snare   & 4.12 & \textbf{1.23} & 1.33 & 1.32 & 1.42 \\
Toms    & 4.78 & 1.45 & 1.45 & \textbf{1.40} & 1.41\\
Hi-Hats & 7.19 & \textbf{3.40} & 3.46 & 4.26 & 4.30 \\
Cymbals & 5.81 & 2.20 & \textbf{1.92} & 2.75 & 3.52 \\
\midrule
\textbf{Overall} & 5.17 & 1.88 & 2.12 & 1.92 & 2.29 \\
\bottomrule
\end{tabular}
\caption{Per-stem mel-spectrogram MSE ($\downarrow$) for drum source separation. Lower values indicate better separation quality.}
\label{tab:MSE_ss_comparison}
\end{table}

\begin{table}[t]
\centering
\small
\setlength{\tabcolsep}{4pt}
\begin{tabular}{l c cccc}
\toprule
& & \multicolumn{4}{c}{\textbf{MSG-LD}} \\
\cmidrule(lr){3-6}
\textbf{Class} & \textbf{LarsNet} & \textbf{Vanilla} & \textbf{(+OB)} & \textbf{(+TB)} & \textbf{(+OB+TB)} \\
\midrule
Kick        & 2.04 & 0.50 & 0.32 & 0.32 & \textbf{0.23} \\
Snare       & 1.20 & 0.13 & 0.10 & 0.12 & \textbf{0.09} \\
Toms        & 0.39 & 0.26 & 0.16 & 0.18 & \textbf{0.13} \\
Hi-Hats     & 2.55 & \textbf{0.35} & 0.38 & 0.50 & 0.48 \\
Cymbals     & 0.92 & 0.83 & \textbf{0.71} & 0.83 & 0.82 \\
\midrule
Mean & 1.42 & 0.41 & \textbf{0.33} & 0.39 & 0.35 \\
\bottomrule
\end{tabular}
\caption{Fréchet Audio Distance (FAD ↓) per-class results. Lower values indicate better audio quality. Mean denotes the unweighted average across the five drum classes.}
\label{tab:fad_results}
\end{table}

\begin{table*}[t]
\centering
\small
\setlength{\tabcolsep}{4pt}
\renewcommand{\arraystretch}{1.08}
\begin{tabular*}{\textwidth}{@{\extracolsep{\fill}} l l *{3}{Y} *{3}{Y} *{3}{Y}@{}}
\toprule
& & \multicolumn{3}{c}{\textbf{ADTOF}} &
    \multicolumn{3}{c}{\textbf{LarsNet}} &
    \multicolumn{3}{c}{\textbf{MSG-LD (+OB)}} \\
\cmidrule(lr){3-5}\cmidrule(lr){6-8}\cmidrule(lr){9-11}
\textbf{Set} & \textbf{Drum Class}
& \textbf{F1} & $P$ & $R$
& \textbf{F1} & $P$ & $R$
& \textbf{F1} & $P$ & $R$ \\
\midrule

\multirow{6}{*}{\textbf{MDB}}
& Kick    & 0.851 & 0.790 & \best{0.923} & 0.899 & 0.893 & 0.906 & \best{0.931} & \best{0.939} & \best{0.923} \\
& Snare   & 0.752 & \best{0.828} & 0.690 & 0.751 & 0.821 & 0.692 & \best{0.760} & 0.791 & \best{0.732} \\
& Toms    & \best{0.520} & \best{0.520} & \best{0.520} & 0.261 & 0.421 & 0.189 & 0.268 & 0.309 & 0.236 \\
& Hi-Hats & \best{0.773} & \best{0.859} & 0.703 & 0.410 & 0.405 & 0.415 & 0.650 & 0.550 & \best{0.794} \\
& Cymbals & \best{0.849} & 0.802 & \best{0.902} & 0.217 & \best{0.873} & 0.124 & 0.481 & 0.663 & 0.377 \\
\cmidrule(lr){2-11}
& \textbf{Overall}
          & \best{0.795} & \best{0.813} & \best{0.778}
          & 0.613 & 0.700 & 0.545
          & 0.707 & 0.709 & 0.704 \\
\midrule

\multirow{6}{*}{\textbf{ENST}}
& Kick    & 0.795 & \best{0.984} & 0.667 & 0.779 & 0.893 & 0.690 & \best{0.821} & 0.941 & \best{0.729} \\
& Snare   & 0.599 & \best{0.932} & 0.441 & 0.606 & 0.787 & 0.493 & \best{0.642} & 0.698 & \best{0.594} \\
& Toms    & \best{0.581} & \best{0.669} & 0.513 & 0.348 & 0.231 & \best{0.698} & 0.491 & 0.415 & 0.602 \\
& Hi-Hats & \best{0.777} & \best{0.936} & \best{0.663} & 0.340 & 0.699 & 0.225 & 0.630 & 0.643 & 0.618 \\
& Cymbals & \best{0.657} & \best{0.788} & 0.563 & 0.349 & 0.243 & \best{0.619} & 0.391 & 0.611 & 0.287 \\
\cmidrule(lr){2-11}
& \textbf{Overall}
          & \best{0.709} & \best{0.908} & 0.582
          & 0.493 & 0.504 & 0.483
          & 0.640 & 0.695 & \best{0.593} \\
\bottomrule
\end{tabular*}
%\caption{Per-class results on \emph{MDB Drums} and \emph{ENST-Drums}: F1/precision/recall. Best result within each dataset/class/metric group is shown in bold.} % For Meta-Reviewer's opinion ----->
% \caption{A Wilcoxon signed-rank test confirms that MSG-LD (+OB) significantly outperforms LarsNet across nearly all drums at the segment level ($p < 0.01$).}
\caption{Drum transcription performance evaluating F1-score, Precision, and Recall under a 50\,ms tolerance window on MDB Drums and ENST-Drums. We compare the proposed separate-and-detect pipeline using MSG-LD (+OB) against an end-to-end baseline ADTOF~\cite{adtof} and a U-Net drum separation baseline LarsNet~\cite{larsnet_stemgmd}. Bold values indicate the best performance for each metric. A Wilcoxon signed-rank test over evaluation segments indicates that MSG-LD (+OB) significantly improves over LarsNet in overall F1 on both datasets ($p<0.01$).}

\label{tab:full_results_combined}
\end{table*}

\begin{table*}[t]
\centering
\small
\setlength{\tabcolsep}{4pt}
\renewcommand{\arraystretch}{1.08}
\begin{tabular*}{\textwidth}{@{\extracolsep{\fill}}%
  l
  *{3}{Y}
  *{3}{Y}
  *{3}{Y}
  *{3}{Y}@{}}
\toprule
& \multicolumn{3}{c}{\textbf{MSG-LD}} &
  \multicolumn{3}{c}{\textbf{MSG-LD (+OB)}} &
  \multicolumn{3}{c}{\textbf{MSG-LD (+TB)}} &
  \multicolumn{3}{c}{\textbf{MSG-LD (+OB+TB)}} \\
\cmidrule(lr){2-4}\cmidrule(lr){5-7}\cmidrule(lr){8-10}\cmidrule(lr){11-13}
\textbf{Set}
& \textbf{F1} & $P$ & $R$
& \textbf{F1} & $P$ & $R$
& \textbf{F1} & $P$ & $R$
& \textbf{F1} & $P$ & $R$ \\
\midrule
\textbf{MDB Overall}
& 0.689 & 0.701 & 0.678
& \textbf{0.707} & \textbf{0.709} & \textbf{0.704}
& 0.698 & 0.706 & 0.691
& 0.690 & \textbf{0.709} & 0.672 \\
\textbf{ENST Overall}
& \textbf{0.645} & \textbf{0.714} & 0.588
& 0.640 & 0.695 & \textbf{0.593}
& 0.626 & 0.677 & 0.582
& 0.636 & 0.703 & 0.580 \\
\midrule
\textbf{Mean overall F1 (2 sets)}
& \multicolumn{3}{l}{0.667}
& \multicolumn{3}{l}{\textbf{0.674}}
& \multicolumn{3}{l}{0.662}
& \multicolumn{3}{l}{0.663} \\
\bottomrule
\end{tabular*}
\caption{Ablation of training-only auxiliary branches on overall transcription performance.
We report overall F1/precision/recall on \emph{MDB Drums} and \emph{ENST-Drums}.
The last row reports the mean overall F1 across the two test sets.}
\label{tab:ablation_overall}
\end{table*}

\section{Experimental Setup}
\label{sec:exp_setup}

\subsection{Datasets and Preprocessing}

The MSG-LD models are trained on a combined corpus of \textbf{StemGMD}~\cite{larsnet_stemgmd} (>1200 hours, synthesized) and \textbf{IDMT-SMT-Drums}~\cite{dittmar2014real} (2.1 hours, real-world acoustic). We evaluate on two unseen datasets: MDB Drums~\cite{mdb} and ENST-Drums~\cite{enst}. Following the MIREX 2017 50/50 split for MDB Drums, the training split is utilized solely for onset-detector hyperparameter tuning, with the remaining 11 tracks reserved for testing. ENST-Drums is evaluated using its official partition.

All audio signals are resampled to 16\,kHz and converted to mono. We compute 64-bin mel spectrograms using a Hann-window STFT with an FFT size of 1024 and a hop size of 160 samples, corresponding to a 10\,ms frame hop. The mel filterbank covers 30\,Hz to 8\,kHz. During separator training, audio files are segmented into 10.24-second clips.

\subsection{Training Details}

All MSG-LD variants are trained on the combined StemGMD and IDMT-SMT-Drums corpus using drum mixtures paired with isolated drum-piece stems. We group the target stems into five drum classes: kick, snare, toms, hi-hats, and cymbals. Unless otherwise specified, the backbone architecture, optimizer, learning-rate schedule, diffusion noise schedule, and sampling procedure follow the original MSG-LD configuration~\cite{msg-ld}. In this work, we keep these settings unchanged and modify the track conditioning to predict five drum-piece stems.

The \textit{Vanilla} model is trained only with the latent diffusion noise-prediction objective. For the \textbf{+OB} variant, MIDI annotations are converted into stem-wise binary onset pianorolls at a 100\,Hz frame rate and used as auxiliary onset targets. For the \textbf{+TB} variant, we pre-compute one 7-D DrumGAN timbre descriptor~\cite{timbre} for each isolated drum stem, resulting in a \(K\times7\) target matrix for each training example, where \(K=5\) corresponds to kick, snare, toms, hi-hats, and cymbals. Each descriptor dimension is linearly normalized to \([0,1]\). The \textbf{+OB+TB} variant enables both auxiliary targets simultaneously. These targets are used only to train the corresponding auxiliary branches defined in Sec.~\ref{sec:method}.

The MSG-LD architecture consists of a trainable U-Net denoiser with 305.17M parameters and a frozen VAE backbone with 128.12M parameters. The auxiliary onset and timbre heads add 0.145M and 0.033M parameters during training, respectively, which represent less than 0.06\% of total trainable parameters and are completely discarded at inference. When an auxiliary branch is disabled, its corresponding loss weight is set to zero; when enabled, the auxiliary loss weights are set to $\lambda_{\mathrm{on}} = 0.05$ with a linear warm-up over the first 2,000 steps, and $\lambda_{\mathrm{tim}} = 0.5$ to balance its initial loss magnitude relative to the diffusion objective at roughly a 1:1 ratio. All variants are trained on a single NVIDIA RTX 6000 Ada Generation GPU with a batch size of 3, using the AdamW optimizer with a base learning rate of $3 \times 10^{-5}$.

\subsection{Model Configurations and Baselines}

We compare four MSG-LD variants: \textit{Vanilla}, \textbf{+OB}, \textbf{+TB}, and \textbf{+OB+TB}. These variants share the same latent diffusion separator and differ only in whether the training-only auxiliary branches are enabled. For baselines, we evaluate ADTOF~\cite{adtof}, a representative strong end-to-end ADT system, and LarsNet~\cite{larsnet_stemgmd}, a strong U-Net-based drum source separation baseline. We use the released checkpoints for both baselines and do not retrain them.

\subsection{Evaluation Metrics}

\noindent \textbf{Drum Source Separation (DSS):}
Following the evaluation setting of latent generative source separation models~\cite{msg-ld}, we do not report SDR or SI-SDR. Since the proposed separator predicts mel-domain latent representations and reconstructs waveforms through a neural vocoder, waveform-domain metrics may entangle separation quality with vocoder reconstruction and phase-related effects. We therefore report per-stem mel-spectrogram mean squared error (\textit{mel-MSE}) as a feature-domain reconstruction measure. Furthermore, to avoid relying solely on a single signal-level metric to evaluate generative distribution fidelity, we report Fréchet Audio Distance (\textit{FAD}). The FAD is computed using VGGish embeddings between the generated stems and their corresponding target stems, and the results are averaged across all drum classes.

\vspace{1mm}
\noindent \textbf{Drum Transcription (ADT):}
Transcription performance is measured using event-based F1-score, Precision (P), and Recall (R) with a 50\,ms tolerance window and one-to-one matching. We report per-class scores for kick, snare, toms, hi-hats, and cymbals. We also report an overall score to summarize system-level performance. For the ablation study, we additionally report the mean overall F1 across MDB Drums and ENST-Drums.

\subsection{Unified Transcription Protocol}

For LarsNet and all MSG-LD variants, separated stems are converted into drum events using the same onset-based post-processing protocol. For each separation front end and drum class, we select the onset detection backend and peak-picking hyperparameters on the MDB training split using Optuna. The search space includes \texttt{madmom} CNN~\cite{madmom}, SuperFlux~\cite{boeck_superflux}, threshold, and minimum inter-onset interval.

After selection, the onset detector configurations are frozen and applied to the MDB test split and to ENST-Drums. ADTOF is evaluated using its official event-decoding procedure. This protocol ensures that each separation front end is evaluated with detector settings selected only on the MDB training split and then frozen for all test results, so the reported differences mainly reflect the quality and onset detectability of the separated stems rather than test-set-specific tuning.

\section{RESULTS \& DISCUSSION}
\label{sec:results}

\subsection{Evaluation of Drum Source Separation}
The reconstruction and distributional audio-quality results are summarized in Table \ref{tab:MSE_ss_comparison} and Table \ref{tab:fad_results}. In terms of physical reconstruction, all MSG-LD variants significantly outperform the LarsNet baseline \cite{larsnet_stemgmd} in mel-spectrogram MSE across all drum classes. While the \textit{Vanilla} model achieves the lowest overall MSE (1.88), we observe that the introduction of auxiliary branches increases the reconstruction error—a trend expected as the model's latent space is regularized toward task-specific features rather than pure signal minimization.

% In terms of generative distribution quality, however, the auxiliary branches offer substantial benefits. As shown in Table \ref{tab:fad_results}, all MSG-LD variants yield much lower Fréchet Audio Distance (FAD) scores compared to LarsNet (1.42). Notably, the \textbf{+OB+TB} variant achieves the highest audio fidelity for foundational classes, reaching the lowest FAD for kick (0.23), snare (0.09), and toms (0.13). This confirms that joint onset and timbre supervision encourages the diffusion model to synthesize high-fidelity textures. For classes like hi-hats, a trade-off emerges: while auxiliaries improves the acoustic fidelity of percussive attacks, they slightly increase the \textbf{hi-hat FAD} compared to the \textbf{+OB} version, suggesting that dense, broadband textures remain sensitive to the conditioning intensity.

In terms of generative distribution quality, however, the auxiliary branches offer substantial benefits. As shown in Table~2, all MSG-LD variants yield much lower Fréchet Audio Distance (FAD) scores compared to LarsNet (1.42). Notably, the +OB+TB variant achieves the lowest FAD for foundational classes, including kick (0.23), snare (0.09), and toms (0.13). This suggests that joint onset and timbre supervision improves embedding-space similarity to the target stems. For classes like hi-hats, a trade-off emerges: while auxiliary supervision may improve attack-related structure, it slightly increases the hi-hat FAD compared to the +OB version, suggesting that dense, broadband textures remain sensitive to the conditioning intensity.

\subsection{Drum Transcription Performance}
Table \ref{tab:full_results_combined} and Table \ref{tab:ablation_overall} present the transcription results using the frozen detection backend. In the ablation study (Table \ref{tab:ablation_overall}), the \textbf{+OB} variant emerges as the most balanced configuration for transcription utility, achieving the highest mean overall F1-score (0.674) across datasets. This variant successfully prioritizes the sharp energy envelopes required for precise onset detection without the spectral blurring occasionally introduced by timbral regression.

When compared against baselines (Table \ref{tab:full_results_combined}), our separate-and-detect pipeline using MSG-LD (+OB) consistently surpasses LarsNet in overall F1-score. Our approach also outperforms the representative end-to-end model (ADTOF \cite{adtof}) in transcribing kick and snare across both datasets. For instance, on MDB, MSG-LD (+OB) reaches an F1-score of 0.931 for kick and 0.760 for snare, compared to 0.851 and 0.752 for ADTOF, respectively. 

To provide deeper visual intuition, Figures \ref{fig:pianoroll} and \ref{fig:waveform_comparison} illustrate the performance on an unseen excerpt. MSG-LD shows strong alignment with the ground truth for foundational classes; it matches LarsNet’s high accuracy on the kick drum while providing a more complete transcription of the snare track. Although transcribing the \textit{toms} class remains difficult, our visualizations reveal specific qualitative advantages. While LarsNet exhibits ringing and cross-stem leakage that can trigger false-positive onsets, MSG-LD (+OB) maintains cleaner silent regions and produces sharper kick and snare transients. This combination of background noise suppression and clearer attacks helps explain its performance advantage over LarsNet in the overall transcription metrics. However, dense hi-hat and cymbal textures remain difficult: both the pianoroll and waveform visualizations show that broadband events can still be blurred or confused across classes.

Transcribing classes with complex sustain or overlapping broadband frequencies remains challenging for separation-based methods. As seen in Figure \ref{fig:pianoroll}, MSG-LD exhibits minor inter-class confusion between hi-hats and cymbals, whereas direct-prediction models like ADTOF excel at capturing cymbals from the mixture. This indicates that broadband overlaps and VAE latent compression are key bottlenecks requiring future optimization. Nevertheless, our pipeline maintains competitive overall performance and excels at extracting the kick and snare. Since these foundational elements constitute the rhythmic backbone of a track, accurately isolating them ensures that our pipeline provides a reliable foundation for both symbolic analysis and practical audio editing workflows.

% \subsection{Limitations and Inference Efficiency}
% Although the separate-and-detect formulation requires a multi-stage inference process, the auxiliary branches (+OB, +TB) are discarded after training, incurring no additional computational overhead at test time. While the iterative sampling of the diffusion model is inherently slower than single-pass end-to-end models like ADTOF (requiring approximately 25 seconds to generate a 10-second sample on an RTX 6000 Ada GPU), it yields interpretable and editable audio stems that end-to-end systems cannot provide. Future work will aim to accelerate the sampling process and mitigate the spectral blurring of broadband frequencies.

\subsection{Limitations and Inference Efficiency}

% While the iterative sampling of the diffusion model is inherently slower during inference than single-pass end-to-end models like ADTOF (requiring approximately 25 seconds to generate a 10-second sample on an NVIDIA RTX 6000 Ada GPU), it yields interpretable and editable audio stems that end-to-end systems cannot provide. Future work will aim to accelerate the sampling process and mitigate the spectral blurring of broadband frequencies.

% While iterative diffusion sampling is slower than single-pass end-to-end models like ADTOF (requiring approximately 25 seconds to generate a 10-second sample on an NVIDIA RTX 6000 Ada GPU), it yields interpretable and editable audio stems. Future work will aim to accelerate sampling, improve high-frequency drum reconstruction, and study robustness to upstream separation artifacts.

The current system has several limitations. First, the 16 kHz setting reduces computation but limits recoverable high-frequency content, which may affect hi-hats and cymbals. Second, iterative diffusion sampling is slower than single-pass end-to-end models like ADTOF, requiring approximately 25 seconds to generate a 10-second sample on an NVIDIA RTX 6000 Ada GPU. Future work will explore faster sampling, higher-resolution latent representations, music-oriented vocoders, and robustness to upstream separation artifacts.

\section{CONCLUSION \& FUTURE WORK}
In this paper, we presented a generative separate-and-detect pipeline for automatic drum transcription. Evaluated on the MDB and ENST datasets, our multi-track latent diffusion separator consistently outperforms the strong U-Net-based baseline (LarsNet) in overall F1-score while remaining competitive with strong end-to-end models. Most notably, the pipeline excels at isolating and transcribing foundational rhythmic elements, especially the kick and snare, while simultaneously providing editable, separated audio stems that direct-prediction systems cannot yield.

% Our ablation analysis revealed a clear decoupling between objective signal reconstruction, generative acoustic quality, and discrete transcription utility. Specifically, auxiliary onset supervision (+OB) provides the strongest transcription-oriented configuration, whereas joint onset and timbre guidance (+OB+TB) yields the highest audio fidelity across foundational drum classes. 

Our ablation analysis suggests a decoupling between feature-domain reconstruction, FAD-based distributional audio quality, and discrete transcription utility. Auxiliary onset supervision (+OB) provides the strongest transcription-oriented configuration, whereas joint onset and timbre guidance (+OB+TB) yields the lowest FAD for several foundational drum classes. Main limitations include the handling of dense, broadband textures like cymbals, as well as the inference latency of iterative diffusion sampling. Future work will explore higher-resolution latent representations, alternative audio codecs, and advanced sampling acceleration techniques for drum transcription and editing.

\section{ACKNOWLEDGMENTS}
The work is supported by grants from Google Asia Pacific, the National Science and Technology Council of Taiwan (NSTC 114-2628-E-002-013-MY3), and the Ministry of Education (MOE) of Taiwan (for Taiwan Centers of Excellence in Artificial Intelligence).

\section{AI Usage Statement}
During the preparation of this work, generative AI technologies were utilized solely as writing and development assistance tools. Specifically, they were used for grammatical editing and text polishing of the manuscript, as well as for modifying minor utility scripts and debugging code during the experimental setup. No generative AI tools were used to design the core methodology, model architecture, or to generate the primary experimental results.

\bibliography{ISMIRtemplate}

% For non BibTeX users:
%\begin{thebibliography}{citations}
% \bibitem{Author:17}
% E.~Author and B.~Authour, ``The title of the conference paper,'' in {\em Proc.
% of the Int. Society for Music Information Retrieval Conf.}, (Suzhou, China),
% pp.~111--117, 2017.
%
% \bibitem{Someone:10}
% A.~Someone, B.~Someone, and C.~Someone, ``The title of the journal paper,''
%  {\em Journal of New Music Research}, vol.~A, pp.~111--222, September 2010.
%
% \bibitem{Person:20}
% O.~Person, {\em Title of the Book}.
% \newblock Montr\'{e}al, Canada: McGill-Queen's University Press, 2021.
%
% \bibitem{Person:09}
% F.~Person and S.~Person, ``Title of a chapter this book,'' in {\em A Book
% Containing Delightful Chapters} (A.~G. Editor, ed.), pp.~58--102, Tokyo,
% Japan: The Publisher, 2009.
%
%\end{thebibliography}

\end{document}